\documentclass[fleqn,usenatbib]{mnras}

\usepackage{newtxtext,newtxmath}

\usepackage[T1]{fontenc}
\usepackage{ae,aecompl}

\usepackage{graphicx}	
\usepackage{amsmath}	
\usepackage{amssymb}	
\usepackage{lipsum}
\usepackage{pdflscape}
\usepackage{afterpage}

\newcommand{\chandra}{\textit{Chandra}}

\newcommand{\swift}{{\it Swift}}

\newcommand{\xmm}{{\it XMM-Newton}}
\newcommand{\rosat}{{\it ROSAT}}
\newcommand{\asca}{{\it ASCA}}
\newcommand{\hubble}{{\it Hubble Space Telescope}}

\title[The UFO in IRAS~13349+2438]{A high velocity component to the complex absorption in IRAS~13349+2438}

\author[M. L. Parker et al.]{M. L. Parker,$^{1}$\thanks{E-mail: mparker@sciops.esa.int}
G. A. Matzeu,$^{1}$
M. Guainazzi,$^{2}$
E. Kalfountzou,$^{1}$
G. Miniutti,$^{3}$\newauthor
M. Santos-Lle\'{o},$^{1}$
and N. Schartel$^{1}$
\\
$^{1}$European Space Agency (ESA), European Space Astronomy Centre (ESAC), E-28691 Villanueva de la Ca\~{n}ada, Madrid, Spain\\
$^{2}$European Space Agency, European Space Research \& Technology Centre (ESTEC), Postbus 299, 2200 AG Noordwijk, The\\ Netherlands\\
$^{3}$Centro de Astrobiolog\'{i}a (CSIC-INTA), Dep. de Astrofis\'{i}ca, E-28691 Villanueva de la Ca\~{n}ada, Madrid, Spain
}

\date{Accepted XXX. Received YYY; in original form ZZZ}

\pubyear{2017}

\begin{document}
\label{firstpage}
\pagerange{\pageref{firstpage}--\pageref{lastpage}}
\maketitle

\begin{abstract}
We present an analysis of \xmm\ spectra of the low-redshift quasar IRAS~13349+2438. The RGS spectrum shows a large number of absorption lines from two zones of warm absorption, with velocities of $\sim-600$~km~s$^{-1}$, as noted by previous authors. Additionally, we find robust evidence from multiple Ly$\alpha$ absorption lines for a previously undiscovered ultra-fast zone of absorption, with an outflow velocity of -0.13$\pm0.01 c$. The warm absorbers and ultra-fast outflow have similar mass outflow rates, around $40\%$ of the Eddington accretion rate, but the kinetic power is dominated by the high velocity gas, which has a power of $\sim4\%$ of the Eddington luminosity.
\end{abstract}




\section{Introduction}

A significant fraction of AGN have been shown to host powerful outflows \citep{Tombesi10}, thought to be launched from the accretion disk and accelerated either by radiation pressure \citep[e.g.][]{Proga00} or magnetic fields \citep{Blandford82}. The fastest, and most powerful by orders of magnitude, are the ultra-fast outflows or UFOs. These are conventionally defined as those outflows with velocities greater than $10^4$~km~s$^{-1}$, and extend up to velocities of $\sim0.4$c \citep[e.g.][]{Reeves18_pds456}. These outflows have significant kinetic power, due to their high velocities, and can potentially drive AGN feedback \citep[see review by][]{Fabian12}.
Several such outflows have been detected in (relatively) high redshift lensed quasars \citep[e.g.][]{Chartas02, Chartas03,Dadina18}, showing that these outflows remain across a wide range of redshifts.

One potential issue raised by the discovery of these outflow is the degeneracy of the scattered emission from the wind with other forms of relativistic emission from AGN, most notably relativistic reflection \citep{Fabian89}. First observed in the AGN MCG--06-30-15 \citep{Tanaka95}, and since observed in many additional AGN and X-ray binaries \citep[XRBS; see review by][]{Middleton16}, reprocessed emission from the inner accretion disk gives a powerful tool to probe the strong gravity regime. However, in sources with powerful relativistic outflows, scattered emission from the wind can potentially produce similar relativistically-broadened line features, due to the broad range of line-of sight velocities found in the wind. While it is generally accepted that relativistic reflection is found in at least some AGN \citep[due to reverberation detections, similarities with X-ray binaries, and microlensing of the accretion disk, e.g.][]{Zoghbi12,Walton12,Chartas17}, it is vitally important for this technique that we understand the potential impact of winds on the measured parameters.

Combining relativistic spectroscopy of UFOs and relativistic reflection is potentially very powerful. The fastest UFOs are thought to be launched within $\sim20$ gravitational radii of the black hole, where the escape velocity is equal to that observed in the outflowing gas. This is the same region that is probed by relativistic reflection, meaning that physical properties of the wind launching region (density, ionization, etc.) can be measured using with reflection spectroscopy and can be used to inform our understanding of UFOs.

IRAS~13349+2438 is a high mass, low redshift \citep[$\sim10^9M_\odot$, $z=0.10853$][]{Lee13} AGN, with significant spectral variability. It is well known for showing several different zones of ionized absorption. The first evidence for a warm absorber in this source came in the form of an edge-like feature at $\sim0.6$~keV in ROSAT and ASCA spectra \citep{Brandt95, Brandt97}, which was initially interpreted as an oxygen edge. However, when the much higher resolution \xmm\ Reflection Grating Spectrometer \citep[RGS;][]{denHerder01} spectrum was analyzed by \citet{Sako01} they determined that this feature was instead due an unresolved transition array (UTA) from Fe \textsc{vii--xii}, the first such astrophysical UTA discovered. In addition, this spectrum revealed a host of absorption lines from Fe, Ne, O, N and C ions, requiring at least two absorption zones along the line of sight, and with outflow velocities of a few hundred km~s$^{-1}$. Later observations with \chandra\ confirmed this \citep{Holczer07,Lee13}, with simultaneous \hubble\ STIS revealing UV absorption lines from gas with a consistent range of velocities \citep{Lee13}.
A relativistic reflection component was identified in the \xmm\ EPIC-pn spectrum by \citet{Longinotti03}, who performed extensive modeling of the broad Fe~K line. Interestingly, they found evidence for a two-component line profile, with a narrow peak at $\sim7$~keV and a broad line from $\sim$5.5--6.5~keV (energies in rest-frame).

In this paper we revisit the \xmm\ spectra of this fascinating source, and identify an ultrafast component to the absorption. We compare continuum models, and explore the impact of having both relativistic reflection and a P-Cygni profile present in the same data.

\section{Observations and data reduction}
We make use of all the available \xmm\ data on IRAS~13349+2438. The observation IDs used are listed in Table~\ref{tab_obsids}. The data were processed using the \xmm\ science analysis software (SAS) version 16.1.0. Data reduction for the EPIC and RGS instruments is described below.

\begin{table}
\centering
\caption{Details of the \xmm\ observations of IRAS~13349+2438. Count rates and exposure times are those of the EPIC-pn. Note that while the count rates differ significantly, this is largely due to changes in hardness so the flux varies only by $\sim\pm30\%$.}
\label{tab_obsids}
\begin{tabular}{l c c r}
\hline
\hline
ObsID		&	Clean exposure	&	Count rate	&	Start Date\\
			&	time (ks)			&	(s$^{-1}$)	\\
\hline
0096010101	&	29.5			&	0.15		&	2000-06-20\\
0402080201	&	25.0			&	0.26		&	2006-07-15\\
0402080301	&	44.8			&	0.25		&	2006-07-16\\
0402080801	&	9.5				&	0.15		&	2006-12-19\\
\hline
\end{tabular}
\end{table}

\subsection{EPIC}
The EPIC-pn and -MOS spectra are reduced using the \textsc{epproc} and \textsc{emproc} tools, respectively, and filtered for background flares. Source spectra are extracted from 40$^{\prime\prime}$ circular regions, centered on the source, avoiding detector edges, and large circular regions on the same chip are used for the background, avoiding the region of the pn detector where the copper background is high. In obsID 0096010101 the MOS2 was operated in timing mode. As this mode is not well calibrated, we do not include this data in our analysis.

Unless otherwise specified, we fit the two MOS spectra independently but group them for clarity in plots using the \textsc{xspec} group command. All EPIC spectra are binned to a signal to noise ratio of 6, after background subtraction, and to oversample the spectral resolution by a minimum factor of 3 \citep{Kaastra16}.

\begin{figure}
\includegraphics[width=\linewidth]{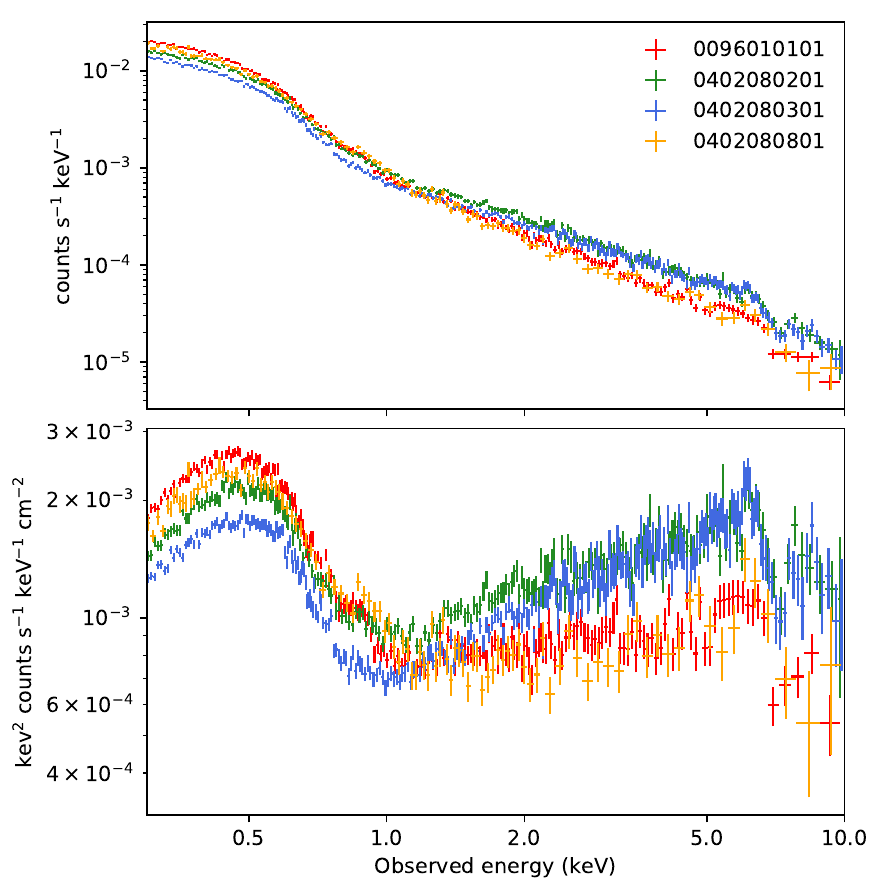}
\caption{Top: EPIC-pn spectra of the four \xmm\ observations. Bottom: As top, but corrected for the effective area and in units of energy flux.}
\label{fig_observations}
\end{figure}

In Fig.~\ref{fig_observations} we show the EPIC-pn spectra of the four \xmm\ observations of IRAS~13349+2438. Interestingly, the variability between observations appears to correspond to changes in spectral hardness, without major changes in the flux. Overall, there is relatively little inter-observation variability, compared to the amount of variability that is routinely integrated over in lower mass sources. For the sake of the analysis presented in this paper, we construct time-averaged spectra for the EPIC-pn and EPIC-MOS using the \textsc{addspec} tool. As shown in \S~\ref{sec_3to10}, the absorption features that are the main subject of this paper do not significantly depend on the overall spectral state of the source. We defer a qualitative discussion of the effect of the spectral variability on our results to \S~\ref{sec_broadband} Unless otherwise specified, we use the time-averaged spectra for all our spectral-fitting.

\subsection{RGS}
The RGS data have been reduced by using the standard \textsc{sas} task \textit{rgsproc} with the most recent calibration files, filtering high background intervals by applying a count rate threshold of 0.2~s$^{-1}$ on the background event files. There were no periods of strong flares, where the background rate exceeds $\sim20\%$ of the total source rate, during these observations.  RGS2 was switched off due to an anomaly for obsIDs 0402080201 and 0402080301. 
As with the EPIC data, we combine all four spectra into a single high-signal spectrum using the \textsc{sas} task \textit{rgscombine}. We fit the RGS spectra from 8--35\AA. When fitting in \textsc{Xspec}, we bin to $\Delta\lambda=0.04\mathring{\mathrm{A}}$ which slightly over-samples the RGS FWHM spectra resolution of $\Delta\lambda=0.06$--$0.08\mathring{\mathrm{A}}$.



\section{Results}
\subsection{Long-term lightcurve}
\begin{figure}
\centering
\includegraphics[width=\linewidth]{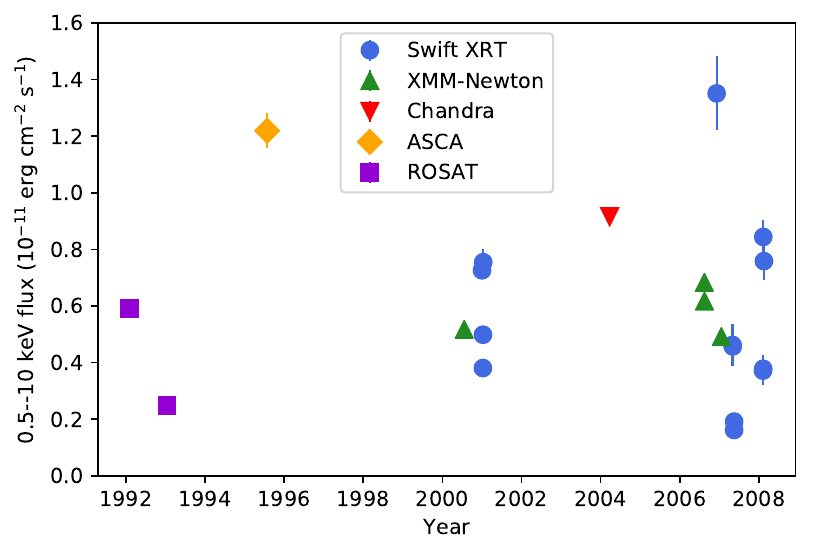}
\caption{Long term lightcurve of IRAS~13349+2438. \swift\ XRT points are from the XRT products generator \citep{Evans07, Evans09}, converted to fluxes by fitting the time-averaged spectrum with a powerlaw plux black-body model and assuming that the count rate is proportional to the flux. \emph{Chandra}, \emph{ASCA} and \emph{ROSAT} fluxes are from \citet{Holczer07,Brandt97} and \citet{Brandt95}, respectively. Where uncertainties are not reported, we assume 5\% error.}
\label{fig_longterm}
\end{figure}

The flux of IRAS~13349+2438 is very variable on long timescales. In Fig.~\ref{fig_longterm} we show a long term lightcurve, with points taken from Neil Gehrels \swift\ Observatory monitoring, the \xmm\ observations reported here, and from literature reports of \chandra , \rosat , and \asca\ observations, to put our observations in context. From this figure it is obvious that the flux has changed by a factor of $\sim10$ in the past, and on timescales much shorter than the 20 year period over which it has been observed. The \xmm\ observations presented here, despite being separated by several years, are similar in flux, and are clustered around the average source flux, so we expect them to be broadly representative of the average source behavior. We note that absorption lines from at least some AGN outflows are flux-dependent \citep{Parker17_nature, Parker18_pds456, Matzeu17}, so by sampling at approximately the same flux we should be seeing features at approximately the same velocity and intensity.

\subsection{3--10~keV spectrum}
\label{sec_3to10}
For this section, as no RGS data is used, we use \textsc{Xspec} \citep{Arnaud96} version 12.9.1p and $\chi^2$ statistics.

\begin{figure}
\includegraphics[width=\linewidth]{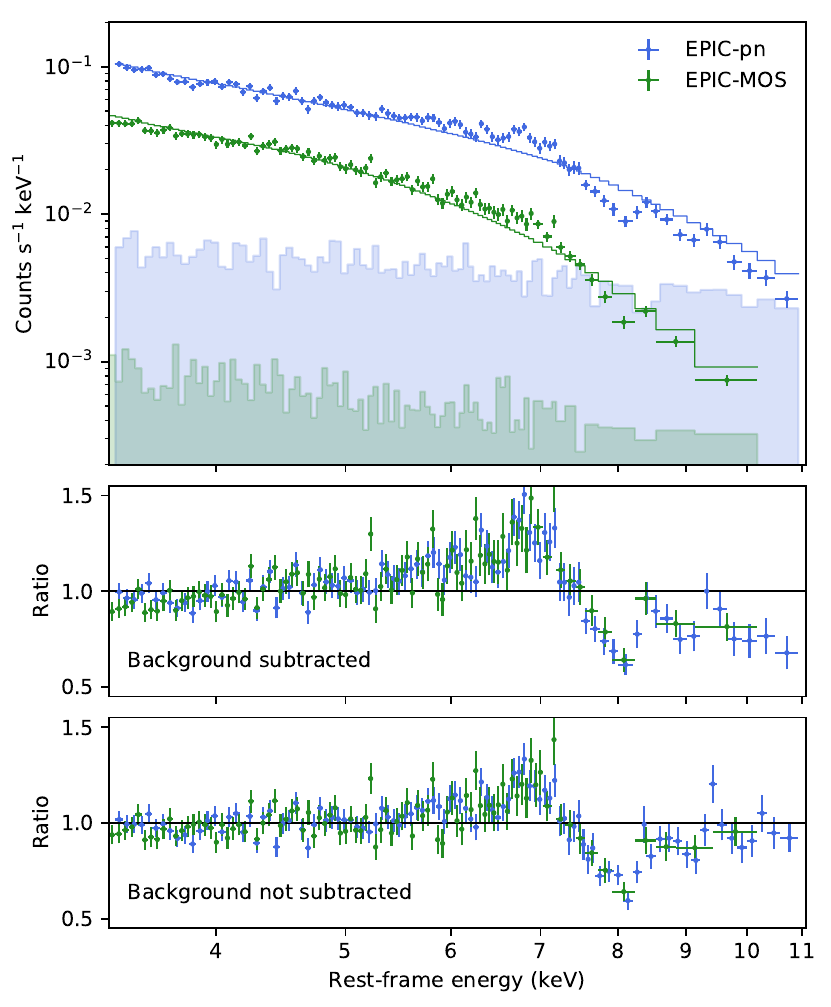}
\caption{Top: 3--10~keV EPIC-pn and MOS spectra, fit with a power-law. Data are rebinned slightly for clarity, and the two MOS spectra are grouped for plotting purposes, but fit separately. Shaded regions show the background spectra. Middle: Residuals to this fit. A broad emission line is visible around 6~keV, and an absorption feature is present at 8~keV. Bottom: As middle, but with no background subtraction.}
\label{fig_background}
\end{figure}

A preliminary investigation of the high energy (3--10~keV) band reveals a residual feature at $\sim8$~keV in the source rest-frame (Fig.~\ref{fig_background}). Because features at these energies can easily be produced by over- or under-subtraction of the instrumental background in the EPIC-pn \citep[which contains strong emission lines from copper and zinc;][]{Carter07} we test the effect of not subtracting the background spectrum. The bottom panels of Fig.~\ref{fig_background} shows the residuals to a power-law fit over this band, with and without the background. The absorption feature is present in both cases, and in the MOS spectrum, which does not have line features in the background at these energies, so we conclude that it is a real absorption line (we discuss the impact of the EPIC background on this feature further in Appendix~\ref{appendix_background}). As no lower energy Fe-K edge or K$\alpha$ line is present, this feature must be due to absorption from the K$\alpha$ line of either Fe\textsc{xxv} or Fe\textsc{xxvi}, the only lines strong enough to produce a single strong absorption feature in this band. If this feature corresponds to Fe\textsc{xxvi}, then it requires a velocity of $\sim0.13c$. If it is due to Fe\textsc{xxv}, then the velocity must be  $\sim0.175c$
An alternative explanation for this feature is rest-frame absorption from the Ni\textsc{xxviii}~K$\alpha$ line at 8.1~keV, which can be produced at higher ionizations when much of the Fe is fully ionized. We explore this in the physical broad-band modeling in section~\ref{sec_broadband}.

\begin{figure}
\includegraphics[width=\linewidth]{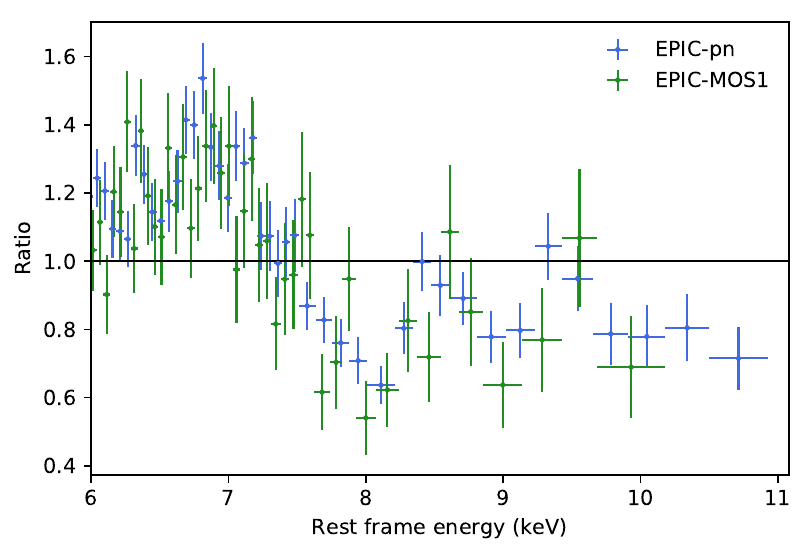}
\caption{Zoom in on the high energy spectral structure, showing the possible secondary absorption line at 9~keV and the flux drop above 9.5~keV. Note that for this figure we only show the MOS1 spectrum, as MOS2 did not take any source exposure during one observation and so is dominated by noise at high energies. Spectra are rebinned slightly in \textsc{xspec} for clarity.}
\label{fig_twolines}
\end{figure}

\begin{figure}
\includegraphics[width=\linewidth]{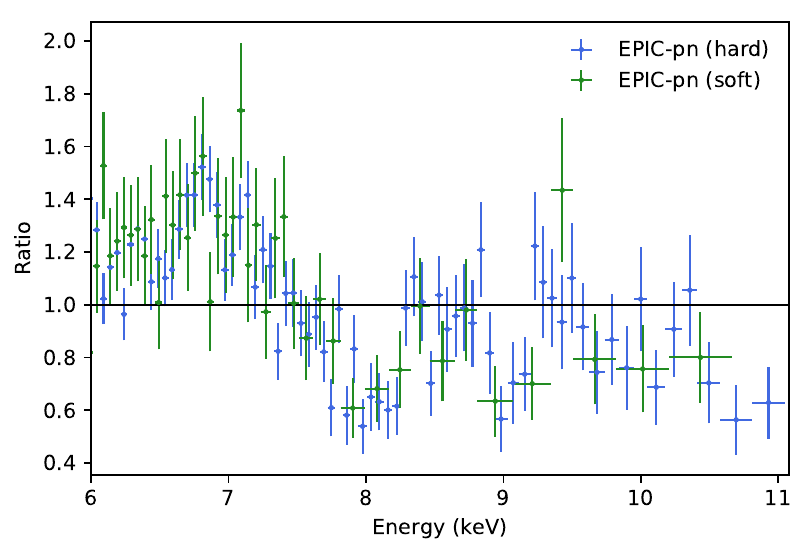}
\caption{High energy spectrum of the EPIC-pn data, grouped into hard and soft spectra (observations 1 and 4 and observations 2 and 3, respectively) and fit with a power-law. Aside from the difference in photon index and flux, the spectral shape is unchanged. Spectra are rebinned slightly in \textsc{xspec} for clarity.}
\label{fig_hardsoft}
\end{figure}

A second line feature is also visible at $\sim9$~keV, although much less significantly. We show a zoom in on this region of the spectrum in Fig.~\ref{fig_twolines}. This feature is seen in both the EPIC-pn and EPIC-MOS1 spectra (MOS2 has 60~ks less exposure than MOS1, and so does not have sufficient signal to meaningfully constrain such a feature). Additionally, we examine the two hard and two soft EPIC-pn spectra separately, finding the same high energy structure in both cases (Fig.~\ref{fig_hardsoft}). If this feature is real, the most likely identification is from Fe\textsc{xxvi} with an outflow velocity of $\sim$0.255$c$. We note that this feature is much more sensitive to the background and high-energy continuum shape than the first line, and coincides with the strong Cu K$\alpha$ line in the EPIC-pn background spectrum (see Appendix~\ref{appendix_background}). We therefore cannot be confident that this line is real without further data.

In Fig.~\ref{fig_hardsoft} we show a comparison of the hard (obs 1 and 4) and soft (obs 2 and 3) EPIC-pn spectra at high energies. From this, it is clear that there is no significant difference in the spectral shape of the absorption feature(s). Based on this, we restrict our analysis of the EPIC data to the time-averaged spectra for the remainder of this work.

\begin{figure*}
\centering
\includegraphics[height=10cm]{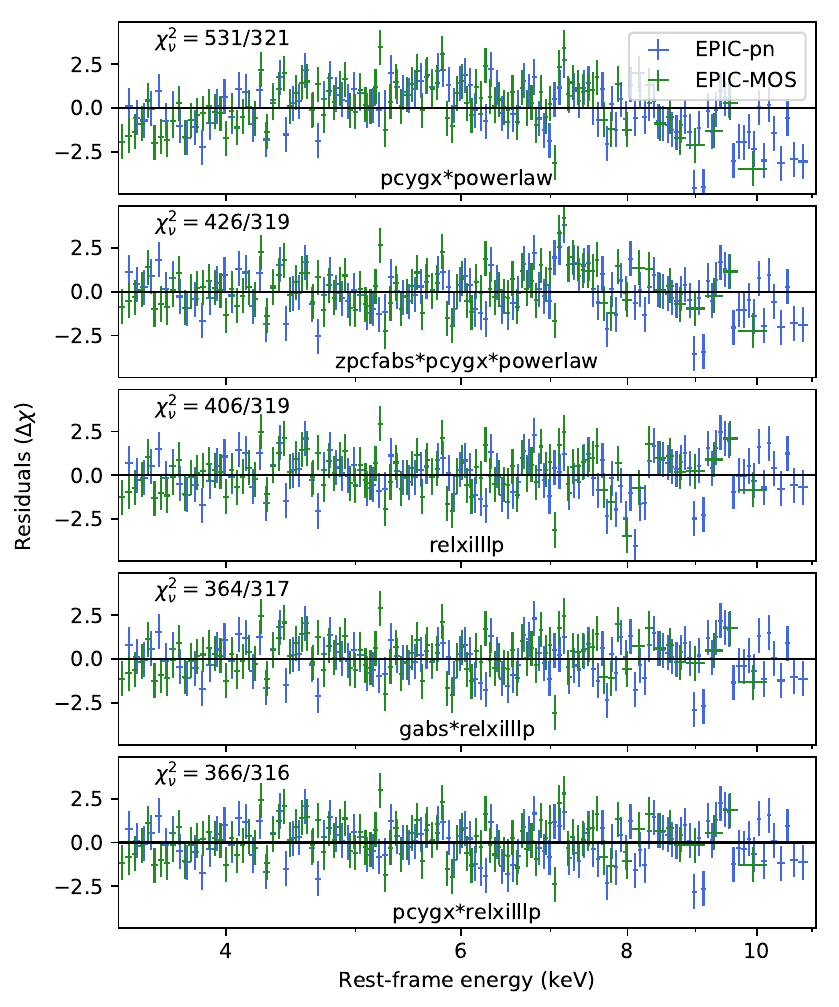}
\includegraphics[height=10cm]{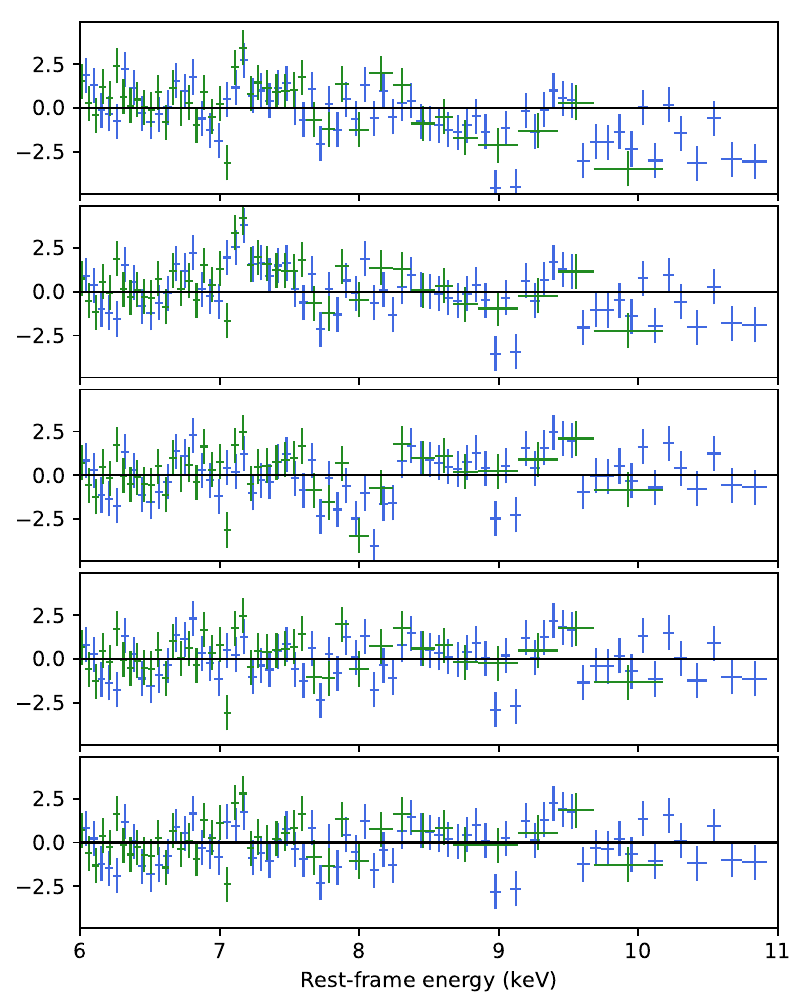}
\caption{Left: Fits to the 3--10 keV band (see text for details). Right: Zoom on the high energy band, showing the absorption line residuals.}
\label{fig_3to10fits}
\end{figure*}

Next, we fit this band with a selection of physically motivated models. As an outflowing wind should produce a broad emission feature from scattered radiation, a P-Cygni profile should naturally be produced in this scenario, at lower energies than the absorption feature \citep{Done07,Nardini15}. To explore this possibility, we use the \emph{pcygx} model \citep{Lamers87,Done07}, which is based on a spherically symmetric outflow, so should be treated as phenomenological. Initially, we apply this model to a power-law (top panel of Fig.~\ref{fig_3to10fits}). This gives a good fit to the absorption feature, but leaves a high degree of spectral curvature across the bandpass, so the fit is not good ($\chi^2_\nu=1.65$). To account for this, we try including a partial covering neutral absorber using \emph{zpcfabs}. This significantly improves the fit ($\chi^2_\nu=1.34$, $\Delta\chi^2=105$, for 2 additional degrees of freedom, 2nd panel of Fig.~\ref{fig_3to10fits}), but leaves a clear residual feature above the iron K edge at 7~keV, where the model predicts a sharp edge not present in the data.

Next, we swap the P-Cygni/partial-covering model for relativistic reflection, using the \textsc{relxilllp} model. This version of the \textsc{relxill} relativistic reflection model \citep{Garcia14} uses a simple lamp-post geometry to parameterize the emissivity profile across the disk. As this is an unphysical geometry, the height parameter should be treated with caution, as it is a proxy for more complex geometrical parameters \citep[see][]{Niedzweicki16}. However, the other relativistic blurring parameters returned by this model are generally consistent with those of models with a broken power-law emissivity profile, and the lamp-post model is less prone to unphysical fits \citep[see discussion in][]{Parker18_tons180}. This model returns a significantly better fit, even without modeling the 8~keV absorption line ($\chi^2_\nu=1.27$, $\Delta\chi^2=20$, for no additional degrees of freedom). With the absorption included, using a simple Gaussian absorption line, the fit improves still further ($\chi^2_\nu=1.14$, $\Delta\chi^2=42$ for 2 additional degrees of freedom, 4th panel of Fig.~\ref{fig_3to10fits}). 

We also try modeling the spectrum with both reflection and a P-Cygni profile simultaneously. This fit is roughly statistically equivalent to the fit with reflection and a Gaussian. Interestingly, this does not strongly affect either the parameters of the reflection or P-Cygni models, which are generally consistent with the parameters found when the other model is not included. This suggests that the fit parameters of the P-Cygni model are constrained primarily by the absorption component, rather than the scattered emission. Using both models simultaneously does moderately lower the iron abundance found with the reflection model, which is otherwise very high. We find $A_\mathrm{Fe}=5\pm2$ with this model, and $A_\mathrm{Fe}>5$ with just reflection and a Gaussian line, although we note that iron abundances (and reflection fractions, which are degenerate with the iron abundance) from a narrow band fit are not reliable.

\begin{table}
\centering
\caption{Best fit model parameters for the 3--10~keV fits.}
\label{tab_3to10pars}
\begin{tabular}{l c c l}
\hline
Component & Parameter & Value & Description\\
\hline
\textsc{gabs}	&	$E$		&	$8.00_{-0.02}^{+0.03}$	& Line energy (keV)\\
				&	$\sigma$	&	$0.1^{*}$	& Line width (keV)\\
                &	EW				&	$0.58^{+0.06}_{-0.04}$	& Equivalent \\
               	&	&	& width (keV)\\
                \\
\textsc{relxilllp}&	$h$				&	$4_{-1}^{+2}$			& Source height\\
				&	$a$				&	$0.93_{-0.02}^{+0.03}$	& Spin\\
                &	$i$				&	$41_{-1}^{+2}$			& Inclination\\
                &	&	& (degrees)\\
                &	$\Gamma$		&	$2.00\pm0.03$			& Photon index\\
                &	$\log(\xi)$ 	&	$3.10\pm0.07$	& Ionization\\
                &	&	&	(erg~s~cm$^{-2}$)\\
                &	$A_\mathrm{Fe}$	&	$>5$					& Iron abundance\\
                &	$R$				&	$6\pm4$					& Reflection \\
                &	&	&fraction\\
                \\
\textsc{constant}& 	$C_\mathrm{pn}$	&	$1^*$			& PN constant\\
				&	$C_\mathrm{MOS1}$&	$1.03\pm0.01$	& MOS1 constant\\
				&	$C_\mathrm{MOS2}$&	$1.13\pm0.01^*$	& MOS2 constant\\
                \\
$\chi^2$/dof	&	&	364/317\\
\hline
\end{tabular}
$^*$The large offset for MOS2 is not due to calibration issues, it is caused by our exclusion of the MOS2 data from obs ID 0096010101, where the source flux was lower than average. 
\end{table}

For brevity, we report only the best-fit parameters for the \textsc{relxilllp} model with a single Gaussian absorption line in Table~\ref{tab_3to10pars}, as this is the simplest model that well describes the continuum, and our main focus in this work is the absorption. For the remainder of this work we will focus on the reflection interpretation of the continuum, but note that other interpretations are possible and cannot be ruled out without a detailed high-energy spectrum. We note that in all of our fits there are residuals at high energies. While we do not include an absorption line from the potential secondary UFO zone, this cannot be the explanation for the features, which appear more like a 9.5~keV emission line than a 9~keV absorption feature. A genuine emission feature at this energy seems highly unlikely, and there are no strong background lines that could be causing this feature, which is present in both the pn and MOS data. It could potentially be caused by additional absorption above 10~keV, an underestimation of the high-energy flux, or a calibration effect.

\subsection{RGS spectra}
Next, we consider the RGS spectrum, aiming to build a good understanding of the warm absorption in the source before fitting the broad-band spectrum. To fit the RGS data in detail, we use the \textsc{spex} spectral fitting package \citep{Kaastra96}, which has been developed specifically for fitting high-resolution X-ray absorption spectra from gratings. We use C-statistics, and dynamically bin the data by a factor of 4 across the bandpass. We use a simple black-body plus power-law continuum, with two \textsc{xabs} photoionized absorption zones to describe the warm absorption \citep[as found by][]{Sako01}. We also include low temperature Galactic absorption, using either the simple \textsc{absm} model \citep{Morrison83}, which accounts for only absorption edges, or the \textsc{hot} model, which is a full collisionally ionized plasma model, including both lines and edges. In both cases, we find that the column drops to zero, and we note that the Galactic column is low in the direction of the source \citep[$\sim10^{20}$~cm$^{-2}$,][]{Kalberla05}. Fixing the column to the predicted value of $10^{20}$~cm$^{-2}$ does not make any statistically significant difference to the fit, and we conclude that Galactic absorption has a negligible effect on the RGS spectrum.

\begin{figure*}
\centering
\includegraphics[width=21cm, angle=90]{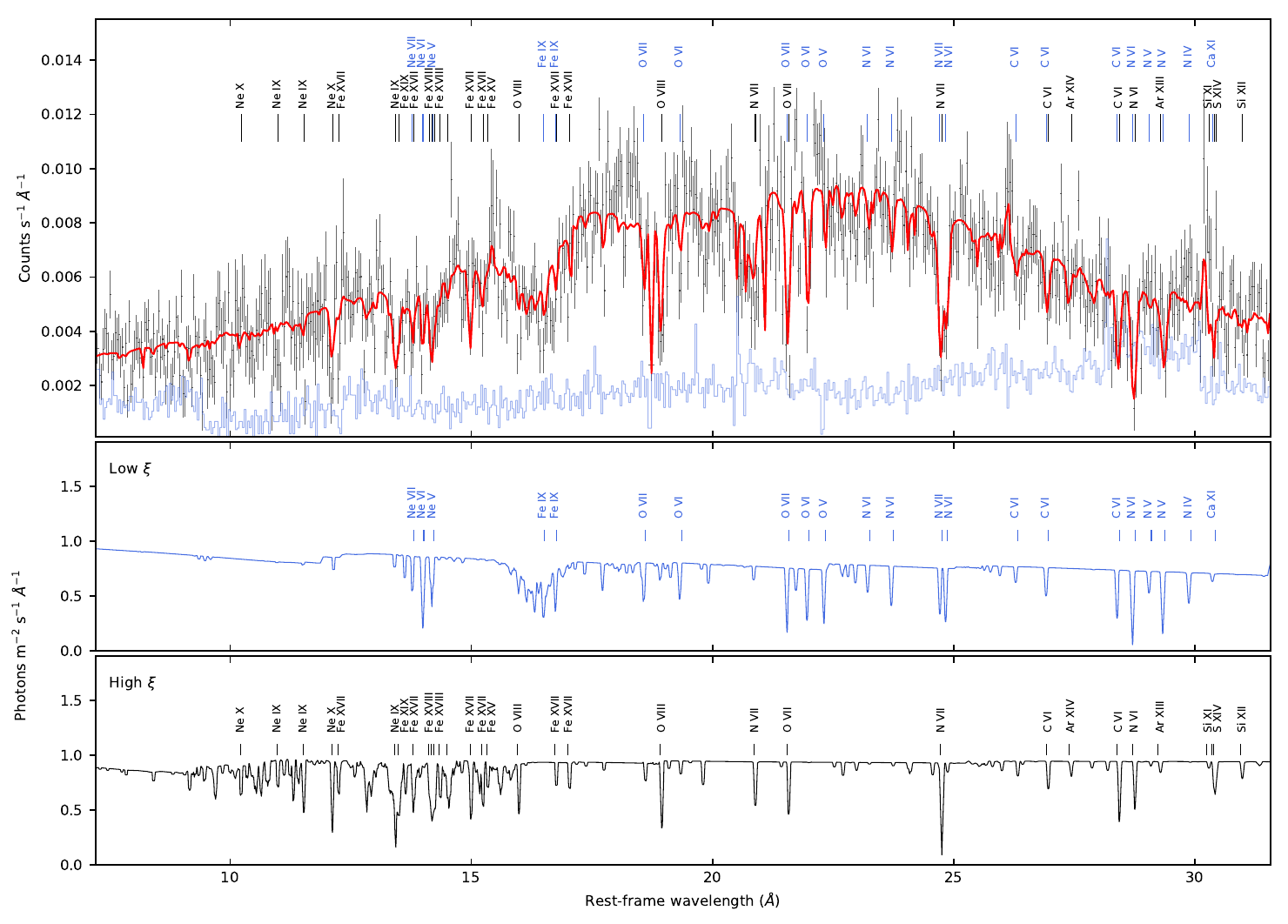}
\caption{Best fit model for the RGS spectrum. We use a black-body plus powerlaw continuum, and two \textsc{xabs} zones. The top panel shows the data and combined model, with the strongest features from both zones labeled. The higher ionization, lower velocity zone is labeled in black, and the lower ionization, higher velocity zone is in blue. The stepped line shows the background spectrum. The second and third panels show the absorption spectra of the low and high ionization zones, respectively.}
\label{fig_rgs_spectrum}
\end{figure*}

The two photoionized absorption zones give a good description of the line features found in the spectrum (shown in Fig.~\ref{fig_rgs_spectrum}). The lower ionization zone produces the Fe UTA from Fe\textsc{vii--xii}, with the strongest features from Fe\textsc{ix}, along a series of O\textsc{v--vii}, lines, Ne\textsc{v--vi} around 14\AA, and N\textsc{iv--vii} at long wavelengths. The higher ionization zone produces the O\textsc{viii} Ly$\alpha$ and $\beta$ lines, most of the N\textsc{vii} Ly$\alpha$ and $\beta$ lines, and a forest of Ne\textsc{ix--x} and Fe\textsc{xv--xix} lines at short wavelengths. Both zones produce equivalently strong C\textsc{vi} lines.

Having established a basic fit, we free the elemental abundances of the strongest ions (C, N, O, Ne and Fe), tying them between the two zones. This significantly improves the fit ($\Delta$ C-Stat$\sim$90, for 5 additional degrees of freedom), with the main driver being a requirement that O is under-abundant with respect to the other elements, for the low ionization zone. Based on this, we fix the other elements to solar values, and re-fit. This gives a comparable overall fit, and an O abundance of 0.2. This fit is shown in Fig.~\ref{fig_rgs_spectrum}

\begin{table}
\caption{Best-fit RGS parameters. All errors are 1$\sigma$.}
\label{table_RGS_pars}
\begin{tabular}{l l c l}
\hline
Component & Parameter & Value & Description/unit\\
\hline
Powerlaw 	&	Norm	&	109$\pm9$	&	$10^{44}$~s$^{-1}$~keV$^{-1}$\\
			&	Gamma	&	1.5$\pm$0.2	&	photon index\\
            \\
Black-body	&	Norm	&	1.5$^{+0.2}_{-0.1}\times10^{-2}$	&	area $(10^{16}$~m$^{2})$\\
			&	T		&	0.098$\pm$0.002	&	keV	\\
            \\
\textsc{xabs}$_1$	&	n$_\mathrm{H}$	&	$4.7_{-0.7}^{+0.8}\times10^{21}$	&	column density\\
			&	&	&	(cm$^{-2}$)\\
			&	log($\xi$)	&	2.40$_{-0.05}^{+0.06}$	&	ionization \\						&	&	&	(erg~s~cm$^{-2}$)\\
            &	$v_\mathrm{RMS}$	&	361$_{-75}^{+124}$	& km~s$^{-1}$\\
            &	$v_\mathrm{out}$	&	-611$_{-96}^{+87}$	& outflow velocity\\ 				&	&	&	(km~s$^{-1}$)\\
            &	$A_\mathrm{O}$	& 0.20$_{-0.04}^{+0.05}$	&	O abundance\\
            \\
\textsc{xabs}$_2$	&	n$_\mathrm{H}$	&	$2.0\pm0.2\times10^{21}$	&	column density\\
			&	&	&	(cm$^{-2}$)\\
			&	log($\xi$)	&	0.85$\pm0.05$	&	ionization \\								&	&	&	(erg~s~cm$^{-2}$)\\
            &	$v_\mathrm{RMS}$	&	267$_{-41}^{+43}$	& km~s$^{-1}$\\
            &	$v_\mathrm{out}$	&	-572$_{-58}^{+67}$	& outflow velocity\\ 				&	&	&	(km~s$^{-1}$)\\
            &	$A_\mathrm{O}$	& 0.20$^*$	&	O abundance\\
            \\
C-stat/dof	&	&	863/650\\
\hline
\end{tabular}
$^*$The oxygen abundance is tied between zones 1 and 2. 
\end{table}


The primary UFO detected in the EPIC data has a velocity of -0.13c. At this ionization, the most likely feature to be present in the RGS data is the O\textsc{viii} Ly$\alpha$ line, which should be around $\sim$16.5\AA. Unfortunately, this coincides with the Fe UTA, so an unambiguous detection of such a line is not possible.

\subsection{EPIC+RGS fit}
\label{sec_broadband}

Having fit the high energy EPIC spectrum and the low energy RGS spectrum, we next attempt a joint fit to the full energy range. As complex models such as \textsc{relxill} cannot be easily transferred to \textsc{spex}, we use \textsc{xspec} for this section. Below 0.5~keV there is a disagreement between the EPIC-pn and MOS spectra, so we restrict these instruments to the 0.5--10~keV band, and fit the RGS from 8--35~\AA ($\sim0.35$--1.55~keV), as before.

We use \textsc{relxilllp} for the continuum, with a phenomenological black body for the soft excess. We model the ionized absorption present in the spectrum with photoionized grids made using the \textsc{xstar} code v2.2 \citep{Bautista01}. The absorption grids were generated in form of multiplicative tables, each computed over the $0.1$--$20\,\rm keV$ band, with $N=10000$ spectral bins, where the $1$--$1000\,\rm Ryd$ photoionizing continuum was assumed to be a power-law with $\Gamma=2$. We generate two grids, one with velocity broadening of $v_{\rm turb}=300\,\rm km\,s^{-1}$ and one with $v_{\rm turb}=1000\,\rm km\,s^{-1}$ for the warm absorption and UFO, respectively, and assume an electron density of $n_{e}=10^{12}\,\rm cm^{-3}$ for both. Elemental abundances are assumed to be solar \citep{Grevesse98}, except for Ni, which is set to zero by default in \textsc{xstar}. The grids cover the $10^{20}\,\rm cm^{-2}<N_{\rm H}<10^{23}\,\rm cm^{-2}$ and $10^{22}\,\rm cm^{-2}<N_{\rm H}<10^{24}\,\rm cm^{-2}$ ranges in $20$ steps of $\Delta N_{\rm H}=2.5\times10^{21}$ for column density and $-1<\log(\xi/\rm erg\,cm\,s^{-1})<4$ and $3<\log(\xi/\rm erg\,cm\,s^{-1})<5$ ranges in $20$ steps of $\Delta(\log\xi=0.1\,\rm erg\,cm\,s^{-1})$ for ionization.

Our baseline model is therefore \textsc{TBnew $\times$ xstar$_1$ $\times$ xstar$_2$ $\times$ (relxill + blackbody)}. This gives a reasonable fit to the combined spectrum (C-stat/dof=1983/1270), but several features remain.
There is a broad dip around 1.2~keV (rest frame), which is not accounted for by the warm absorption models and is not resolved into a narrow feature or features by the RGS. We model this with a broad ($\sigma\sim0.1$~keV) Gaussian absorption line, which accounts for all the residuals, and improves the fit to C-stat/dof=1642/1267 ($\Delta$C-stat=341, for 3 degrees of freedom). The origin of this feature is not immediately obvious, but we note that a similar broad dip is present in the spectrum of IRAS~13224-3809 \citep{Parker17_nature}, which is produced mainly by the Ne\textsc{x} Ly$\alpha$ line from the UFO, with some contribution from Fe\textsc{xxiv}. At the same velocity of the Fe\textsc{xxvi} line, the Ne\textsc{x} Ly$\alpha$ should be present at $\sim1.06$~keV (observed frame). Alternatively, we note that this dip is very close to the point where the soft excess joins the continuum. For simplicity, we have used a phenomenological black-body model for the soft excess, and it is possible that a more complex soft excess (e.g. cool Comptonization, double relativistic reflection, etc.) could account for the observed spectral complexity. Similar features have been observed before: \citet{Leighly97} find $\sim$1~keV dips in 3 narrow-line Seyfert 1 (NLS1) AGN, which they interpret as blueshifted O\textsc{vii} and O\textsc{viii} edges from a relativistic outflow. The main UFO zone is not high enough velocity for the O\textsc{viii} to appear at 1.2~keV, but the possible second zone is. However, it is unlikely that the ionization of this gas (if it exists) is low enough to permit a significant O edge. \citet{Nicastro99} show that such a broad absorption feature can be produced by a large number of Fe~L lines at 1~keV with no significant blueshift when a steep ionizing spectrum is considered. In this case, as the absorption is at 1.2~keV, this interpretation would likely still require a significant blueshift, but could be produced by colder clumps of higher density gas embedded in the UFO.

\begin{figure}
\centering
\includegraphics[width=0.9\linewidth]{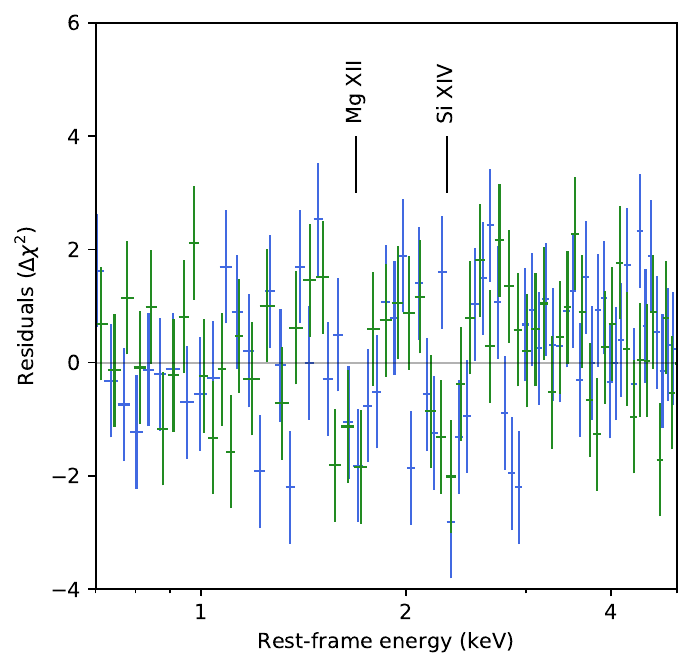}
\caption{Mg and Si lines at $v\sim-0.13c$ found in the EPIC spectra. The vertical lines correspond to the predicted energies of the Ly$\alpha$ lines, based on the velocity shift of the Fe~\textsc{xxvi} line at high energies. The rest-frame energies of these features are $1.68\pm0.02$~keV and 2.25$\pm0.03$~keV, based on fitting Gaussian lines.}
\label{fig_MgandSi}
\end{figure}

After accounting for this feature, there are still two weaker absorption lines present just above and below 2~keV present in the EPIC data, shown in Fig.~\ref{fig_MgandSi}. These coincide well with the predicted energies of the Mg\textsc{xii} and Si\textsc{xiv} Ly$\alpha$ lines from the UFO, at the same velocity (-0.13$c$) as the Fe\textsc{xxvi} line.
Adding Gaussian absorption lines with $\sigma=0.1$~keV and energies fixed at the predicted energies of Mg\textsc{xii} Ly$\alpha$ and Si\textsc{xiv} Ly$\alpha$ for a 0.13$c$ UFO gives improvements of $\Delta\chi^2=15$ and $\Delta\chi^2=38$, respectively, for one additional degree of freedom each. This corresponds to formal significances of $3.9\sigma$ and $6.2\sigma$.
We calculate more accurate significances using a Monte-Carlo approach. We simulate 10,000 spectra, based on the best fit model, with no UFO absorption. For each spectrum, we add a Gaussian absorption line at each of the three intermediate Ly$\alpha$ lines that could realistically be detected in this data (Mg\textsc{xii}, Si\textsc{xiv}, and S\textsc{xvi}), for the two predicted UFO velocities (corresponding to the Fe\textsc{xxv} and Fe \textsc{xxvi} solutions), and record the $\Delta\chi^2$ for each line. This gives the probability of a false positive detection of Mg\textsc{xii} at the observed significance or higher (3 spectra, $P=0.0003$), Si\textsc{xiv} at the observed significance or higher (0 spectra, $P<0.0001$), both lines at the observed significance (0 spectra, $P<0.0001$), and any two lines (including S) with $\Delta\chi^2\geq15$ (0 spectra, $P<0.0001$). By the same method, the probability of a false detection of a S\textsc{xvi} line at $>3\sigma$ is 0.001.

The RGS spectrum is consistent with the lower energy feature, but has little sensitivity. The higher energy feature is outside the bandpass entirely. The calibration of the EPIC instruments is generally extremely good at intermediate energies, and the agreement between the pn and MOS detectors is excellent. The closest instrumental edge is at 2.25 keV (2.5keV in the source rest frame), and does not coincide with either absorption feature. The most likely edge effect is a false emission line above the edge, due to pileup. These observations are below the pileup limit, and no such feature is present.
Neither the \citet{Holczer07}, \citet{Lee13}, or our own warm absorber models predict absorption lines at these energies (in fact, a broad feature is visible at 1.7~keV/7.4\AA in Fig.~3 of \citeauthor{Holczer07} which could be the Mg line), whereas they are consistent in both ionization range and velocity with the UFO Fe line. Given that UFO lines are typically broader than warm absorber lines, the higher sensitivity, lower resolution EPIC instruments are more likely to detect them than the \chandra\ gratings, which is likely why the lines were not found by these authors.

To account for this UFO absorption, we add an additional absorption zone, modeled with an \emph{xstar} grid with a higher velocity broadening of 1000~km~s$^{-1}$ but otherwise the same. This zone successfully fits the Fe, Si and Mg lines, but does not account for 1~keV feature, which still requires an additional Gaussian line to fit. The additional zone (\textsc{xstar}$_3$) improves the fit by $\Delta$C-stat=115, for 3 degrees of freedom. Our final best-fit parameters are given in Table~\ref{table_broadband_pars}. Our final model is \textsc{TBnew $\times$ Gauss $\times$ xstar$_1$ $\times$ xstar$_2$ $\times$ xstar$_3$ $\times$ (relxill + blackbody)}.
The \emph{xstar} default settings have the Ni abundance set to zero, as it is not well calibrated. However, the high energy absorption line is very close to the energy of the Ni~K$\alpha$ absorption line, and could potentially be produced by non-outflowing Ni. To test this, we construct an additional \emph{xstar} grid with a free Ni abundance, and fit to the data with the gas fixed at the rest frame of the source. The ionization is significantly higher ($\log(\xi)>4.8$ erg~cm~s$^{-1}$), so that no Fe~\textsc{xxvi} line is produced. This gives a good description of the high energy absorption line, although it prefers a high but poorly constrained Ni abundance ($A_\mathrm{Ni}=34^{+59}_{-29}$).
Overall, the total fit statistic is worse by $\Delta$~C-stat of $\sim30$, primarily because the ionization is too high to produce a Si\textsc{xiv} line at 2~keV. We conclude that Fe\textsc{xxvi} absorption is the most plausible solution for the high energy absorption line, as it gives a consistent explanation for this feature and the two lower energy lines.

\begin{table}
\caption{Best-fit broadband parameters. All errors are 1$\sigma$.}
\label{table_broadband_pars}
\begin{tabular}{l l c l}
\hline
Component & Parameter & Value & Description/unit\\
\hline
\textsc{gabs}	&	$E$			&	$1.22\pm0.01$	& Rest-frame \\
			&	&	&	energy (keV)\\
            &	$\sigma$	&	$0.12\pm0.01$	& Line width\\
            &	&	&	(keV)\\
            \\
\textsc{xstar}$_1$	&	$n_\mathrm{H}$	&	$4.1_{-0.2}^{+0.4}\times10^{21}$	&	Column density \\
			&	&	&(cm$^{-2}$)\\
            &	log($\xi$)	&	2.27$_{-0.02}^{+0.03}$	&	ionization \\						&	&	&	(erg~cm~s$^{-1}$)\\
            &	$v_\mathrm{out}$	&	-578$_{-132}^{+127}$	& outflow velocity\\ 			 &	 &	 &	 (km~s$^{-1}$)\\
            \\
\textsc{xstar}$_2$	&	$n_\mathrm{H}$	&	$2.15_{-0.08}^{+0.06}\times10^{21}$	&	Column density \\
			&	&	&(cm$^{-2}$)\\
            &	log($\xi$)	&	1.11$_{-0.02}^{+0.03}$	&	ionization \\						&	&	&	(erg~cm~s$^{-1}$)\\
            &	$v_\mathrm{out}$	&	$-524_{-72}^{+61}$	& outflow velocity\\ 			 	&	 &	 &	 (km~s$^{-1}$)\\
            \\
\textsc{xstar}$_3$	&	$n_\mathrm{H}$	&	$5_{-2}^{+3}\times10^{23}$	&	Column density \\
			&	&	&(cm$^{-2}$)\\
            &	log($\xi$)	&	4.27$_{-0.15}^{+0.07}$	&	ionization \\						&	&	&	(erg~cm~s$^{-1}$)\\
            &	$v_\mathrm{out}$	&	-37624$_{-553}^{+136}$	& outflow velocity\\ 			 &	 &	 &	 (km~s$^{-1}$)\\
            \\
Black-body	&	Norm	&	1.5$^{+0.2}_{-0.1}\times10^{-2}$	& (10$^{39}$~erg~s$^{-1}$\\
			&	&	& / 10~kpc$^2$)\\
			&	T		&	0.093$_{-0.001}^{+0.003}$	&	keV	\\
			\\
\textsc{relxilllp}&	$h$				&	$<5$			& Source height (ISCO)\\
				&	$a$				&	$>0.97$	& Spin\\
                &	$i$				&	$48\pm1$			& Inclination\\
                &	&	& (degrees)\\
                &	$\Gamma$		&	$1.96_{-0.01}^{+0.02}$			& Photon index\\
                &	$\log(\xi)$ 			&	$1.2_{-0.2}^{+0.1}$	& Ionization\\
                &	&	&	(erg~s~cm$^{-2}$)\\
                &	$A_\mathrm{Fe}$	&	$>8$					& Iron abundance\\
                &	$R$				&	$7.2_{-0.3}^{+0.6}$					& Reflection \\
                &	&	&fraction\\
                \\
\textsc{constant}& 	$C_\mathrm{pn}$	&	$1^*$			& PN constant\\
				&	$C_\mathrm{RGS}$&	$1.03\pm0.01$	& MOS1 constant\\
				&	$C_\mathrm{MOS1}$&	$1.02\pm0.01$	& MOS1 constant\\
				&	$C_\mathrm{MOS2}$&	$1.12\pm0.01$	& MOS2 constant\\
                \\
C-stat/dof	&	&	1527/1264 \\
\hline
\end{tabular}
\end{table}

\begin{figure*}
\centering
\includegraphics[width=0.9\linewidth]{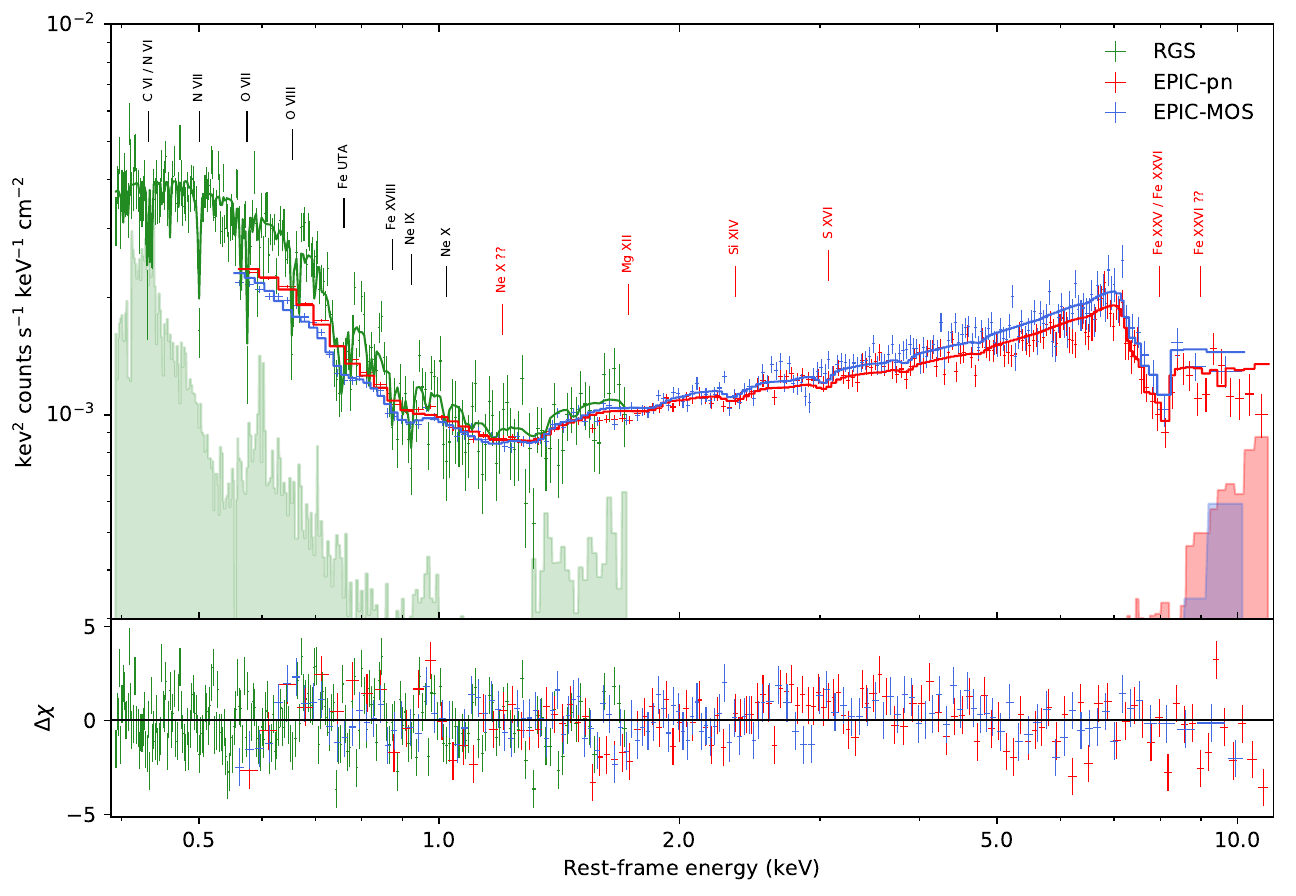}
\caption{Full band EPIC and RGS spectrum, corrected for the instrument effective area but not unfolded. Black vertical lines label the main warm absorber absorption features, and red lines the detected and potential UFO features (we also label the S\textsc{xvi} line, which we do not find in the data, but is the next strongest line in the model). For simplicity, both warm absorbers are labeled together, and only the strongest lines are identified. The shaded region shows the background for each instrument. For clarity, the data are rebinned slightly in excess of that described in the text.}
\label{fig_bestfit_spectrum}
\end{figure*}

\begin{figure}
\centering
\includegraphics[width=\linewidth]{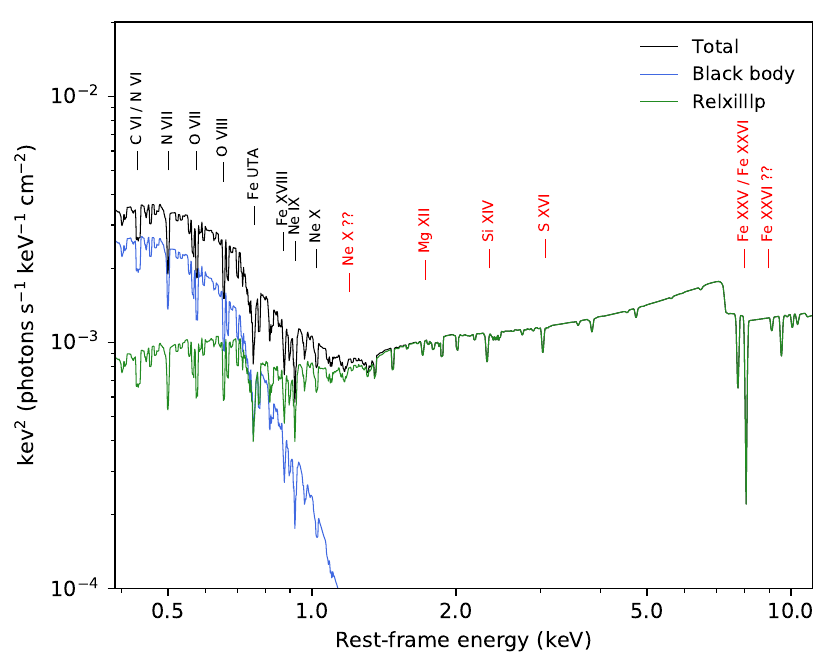}
\caption{Best fit model for the full-band spectrum. Absorption lines are marked as in Fig.~\ref{fig_bestfit_spectrum}}.
\label{fig_bestfit_model}
\end{figure}

The full-band best-fit is shown in Fig.~\ref{fig_bestfit_spectrum}, and the model in Fig.~\ref{fig_bestfit_model}. For Fig.~\ref{fig_bestfit_spectrum}, we correct the data for the effective area using the \texttt{setplot area} command in \textsc{xspec}. This divides the count rate in each channel by the effective area of the channel, but does not attempt to correct for the instrumental resolution. This gives a clear picture of the spectral shape and allows easy comparison of the different instrument spectra, but unlike unfolded spectra the position of the data points is not model-dependent, so it does not exaggerate narrow model features. We note that it is possible the pivoting seen in Fig.~\ref{fig_observations} is affecting the continuum parameters of this fit (in particular, the 1~keV dip could be potentially be produced by this). To explore this possibility, we fit this model to pn spectra of the four observations, fixing all parameters to their best-fit values apart apart from the normalizations of the black-body and \textsc{relxilllp}, the photon index, and the black-body temperature. This gives a reasonable fit to the data ($\chi^2/$dof=521/431=1.2) with no obvious systematic residuals. The 1~keV feature is still strongly required by the data (removing it gives a $\Delta\chi^2$ of $+220$), and we conclude that this feature is genuine and that our parameters of interest are unlikely to be strongly affected by the pivoting.

\section{Discussion}

The long-term lightcurve shows that the source flux increases by at least a factor of $\sim2$--3 relative to the \xmm\ spectra presented here. As there is no evidence for a significant column of neutral absorption, and no major changes in spectral hardness with XRT flux, this suggests that this variability is due to changes in the intrinsic X-ray flux of the source. Given the high mass of this AGN such large variability is unusual, particularly when it cannot be due to eclipsing events. This is presumably due to the high accretion rate of IRAS~13349+2438, possibly combined with changes in the mass-loss rate from the wind.

With the exception of the newly-discovered UFO, our absorption parameters are broadly consistent with past results: the warm absorption is dominated by two main zones, one hot and one cold, with a similar outflow velocity of a few hundred km~s$^{-1}$ \citep{Sako01,Holczer07,Lee13}. The two warm absorption zones in the RGS are consistent with having the same velocity. The most likely explanation for this is that they are in fact part of the same outflow, presumably with cold clumps of higher density material embedded in a hotter, less dense flow as seen in simulations \citep[e.g.][]{Nayakshin07} and found in other AGN \citep[e.g.][]{Reeves18_pg1211}.
We note that the the warm absorption has not changed greatly between the observations presented here. This is not unexpected, given the high mass of the source and the large radius inferred for this material (see below), and it makes the spectra considerably easier to fit.

We also reexamine the broad Fe emission feature identified by \citet{Longinotti03}, and confirm that it is most likely due to relativistic reflection from the inner disk, although we do not find evidence for the double-peaked structure seen by \citeauthor{Longinotti13}. A P-Cygni profile alone cannot reproduce the observed spectral shape, requiring either partial covering absorption or relativistic reflection in addition. As there is no evidence for a substantial column of neutral absorption in this source, we prefer the reflection interpretation. 

The reflection parameters we find are generally consistent with those found in other AGN: high spin, a compact corona, and a moderate inclination. We note, however, that the iron abundance we find is highly super-solar. As discussed in \citet{Parker18_tons180}, this is a frequent problem with reflection modeling, and could be due to several different reasons. One possibility specific to sources with major outflows is overlap with scattered emission from the wind, in the form of a P-Cygni profile.
By combining P-Cygni profiles with relativistic reflection, some of the flux from the broad iron line can be removed, potentially lowering the iron abundance. We explored this possibility in \S~\ref{sec_3to10}, finding that using both models simultaneously moderately lowered $A_\mathrm{Fe}$ but otherwise did not strongly affect the reflection or absorption parameters. This is a potentially promising area for further study with physical wind models. 
Additionally, we note that in high accretion rate sources like IRAS~13349+2438 the thin disk approximation is unlikely to be valid. Thickening of the disk can lead to an enhancement of the red wing of the iron line \citep{Taylor17}, giving higher spin, lower inclination, and lower source height values.

The co-existence of strong relativistic reflection and a UFO in several sources is interesting, as it raises the possibility that the two are linked. In particular, \citet{Gallo11_ufos,Gallo13_ufos} discuss the possibility that the UFO features are produced by a layer on the surface of the accretion disk, where the high velocities occur naturally, and are imprinted on the reflection spectrum. This requires a high reflection fraction, as observed here and in IRAS~13224-3809 \citep{Parker17_nature}, for the absorption features to appear. It also predicts that the reflection parameters and the parameters of the UFO should be linked, in particular that the spin and inclination should strongly affect the velocity of the UFO. We note that the inclination we find in this work ($\sim40$\textdegree) is lower than that of IRAS~13224-3809 ($\sim60$\textdegree), and the velocity of the UFO is correspondingly lower as predicted by this model. An alternative link between UFOs and relativistic reflection is that a strong wind may be required in high accretion rate AGN to keep the inner disk cool and stable, allowing strong relativistic reflection to occur. If we take both UFO and reflection phenomena at face value, then the observed velocity of the outflow should be partially dependent on the viewing angle, which can be measured with reflection. By combining these two measurements, it is therefore possible to constrain the outflow geometry \citep{Parker18_geometry}.

Assuming that the UFO is a genuine outflow, it completely dominates over the warm absorbers in terms of kinetic power. Following \citep{Tombesi13} we calculate minimum and maximum radii for the different outflow zones ($r_\mathrm{min}=2GM_\mathrm{BH}/v_\mathrm{out}^2$, $r_\mathrm{max}=L_\mathrm{ion}/\xi N_\mathrm{H}$), using the ionizing flux from our model (1.5$\times10^{-11}$~erg~s$^{-1}$~cm$^{-2}$), which gives a luminosity of $L_\mathrm{ion}=4\times10^{44}$~erg~s$^{-1}$ (taking $H_0=73$~km~s$^{_1}$~Mpc$^{_1}$, $\Omega_\mathrm{M}=0.27$, $\Omega_\Lambda=0.73$). Assuming that the two warm absorption zones are co-located in a clumpy wind, and taking the stricter constraint on $r_\mathrm{max}$ from the higher ionization zone, the (logarithmic) mean radius of the warm absorber is $1.5\times10^{20}$~cm, or $10^6$ gravitational radii ($R_\mathrm{G}$). For the UFO, this gives a range of $(1.5-3.6)\times10^{16}cm$ (100--240~$R_\mathrm{G}$), and we therefore assume a radius of 150~$R_\mathrm{G}$. From these, we calculate the mass outflow rates, using the equation of \citet{Krongold07} and the same assumptions as \citet{Tombesi13}. We find outflow rates of $5.4\times10^{26}$~g~s$^{-1}$ and $2.2\times10^{26}$~g~s$^{-1}$ for the high and low ionization zones, and $5.8\times10^{26}$~g~s$^{-1}$ for the UFO (0.40,0.16 and 0.42 $\dot{M}_\mathrm{Edd}$, respectively, assuming an efficiency of 0.1). The kinetic power ($\dot{M}v^2/2$) is then $9.8\times10^{41}$~erg~s$^{-1}$, $3.9\times10^{41}$~erg~s$^{-1}$, and $5.1\times10^{45}$~erg~s$^{-1}$, respectively ($8\times10^{-6}\ L_\mathrm{Edd}$, $3\times10^{-6}\ L_\mathrm{Edd}$, $0.04\ L_\mathrm{Edd}$). From this, it is obvious that a large fraction of the accreted matter is lost in the form of large scale outflows over several orders of magnitude in radius. The bolometric luminosity estimated from the source SED by \citet{Lee13} is $\sim3\times10^{46}$~erg~s$^{-1}$, or $\sim0.25$~$L_\mathrm{Edd}$, implying that the majority of the accreted matter is lost in outflows, rather than converted to radiation, although the majority of the energy output is still in the form of radiation. This also implies that the source could easily exceed the Eddington accretion rate before matter is lost in the outflows, although there is a large uncertainty in these values. 

It has been suggested \citep[e.g.][]{Tombesi13} that UFOs and warm absorbers for part of a single large-scale wind, with different ionizations corresponding to different radii. Our results are certainly consistent with those of \citeauthor{Tombesi13}, in that we find low velocity, low ionization absorption at large radii, and high velocity, high ionization absorption at small radii, but in the case of a single large-scale wind structure we would also expect to see some material at intermediate velocities at radii somewhere between 200 and $10^6$~$R_\mathrm{G}$. There is generally a shortage of such material in AGN, which suggests that either the two types of outflow do not have a common origin or the wind is suppressed at intermediate radii by some mechanism. \citet{Laha16} observe different parameter correlations between the UFOs and warm absorbers, and therefore argue that they represent two different populations of outflows, where the UFOs are launched from the inner disk and the warm absorbers are instead photo-evaporated from the torus. \citet{Pounds13} and \citet{King14} consider an alternative explanation, where the warm absorber is produced in the shock where a UFO collides with the shell of gas swept out by AGN radiation, naturally producing a bi-modal distribution of velocities and ionizations. The radii we find for the warm absorbers are consistent with this scenario, although the lower limit on radius is calculated based on the escape velocity of the disk and is therefore not necessarily valid in this interpretation.

\section{Conclusions}
We have analyzed \xmm\ RGS and EPIC spectra of the bright low-redshift quasar IRAS~13349+2438. We have fit detailed absorption models to the spectra, and identify a new ultra-fast zone of absorption. Our key results are summarized here:
\begin{itemize}
\item In the EPIC spectra, we find Fe\textsc{xxvi}, Si\textsc{xiv} and Mg\textsc{xii} absorption lines from a UFO with $v=0.13c$.
\item The majority of the absorption in the time-averaged RGS spectrum is well described by two warm absorption zones, with similar outflow velocities ($\sim600$~km~s$^{-1}$), as found by previous authors. It is likely that these two are co-located, with the difference in ionization being due to clumps in an outflowing wind or shock region.
\item There is a lack of absorption with velocities intermediate between the UFO and warm absorption, implying that the two have different origins, either physically separate or produced by different mechanisms.
\item The mass outflow rate is very high, well in excess of the accretion rate onto the black hole, and a significant amount of kinetic power is released in the UFO, potentially enough to drive AGN feedback.
\item The broad-band continuum is well described by relativistic reflection, with high spin and a moderate inclination. The iron abundance is very high, but this could potentially be mitigated by a contribution from scattered emission from the wind.
\end{itemize}
IRAS~13349+2438 is a well known case-study for ionized absorption in AGN, with multiple absorption zones across a wide range of ionizations. The discovery of this high velocity outflow has greatly expanded our knowledge of this fascinating source.

\section*{Acknowledgements}
We thank the referee, Ehud Behar, for detailed and constructive feedback, which has significantly improved this work.
MLP, GAM and EK are supported by European Space Agency (ESA) Research Fellowships. Based on observations obtained with XMM-Newton, an ESA science mission with instruments and contributions directly funded by ESA Member States and NASA. The authors dedicate this work to the memory of professors Donald Lynden-Bell and \mbox{Yasuo} Tanaka. Without their momentous contributions to the study of black hole astrophysics this paper, and many others like it, would not have been possible.




\bibliographystyle{mnras}
\bibliography{bibliography_iras13349} 


\appendix
\section{EPIC background}

Absorption lines found in EPIC-pn spectra around 8~keV in the observed frame are known to be sensitive to the background treatment, due to the presence of a complex of K$\alpha$ lines from Zn, Cu and Ni in the internal instrumental background \citep[e.g.]{Lumb02}. Over-subtraction of these lines can produce a false absorption feature, leading to the false detection of an outflow where none is present. This is most likely to occur from selection of the background region. The Cu background line is variable across the pn detector, giving a much stronger feature in the outer parts of the detector. Because the observations presented here were taken in small window mode this is less of a risk, as the high-copper region is outside the area exposed in small window mode. However, a careful examination of the background is useful in ensuring that the shape of the line profile is not being affected by over- or under-subtraction of line features.

\begin{figure*}
\centering
\includegraphics[width=\linewidth]{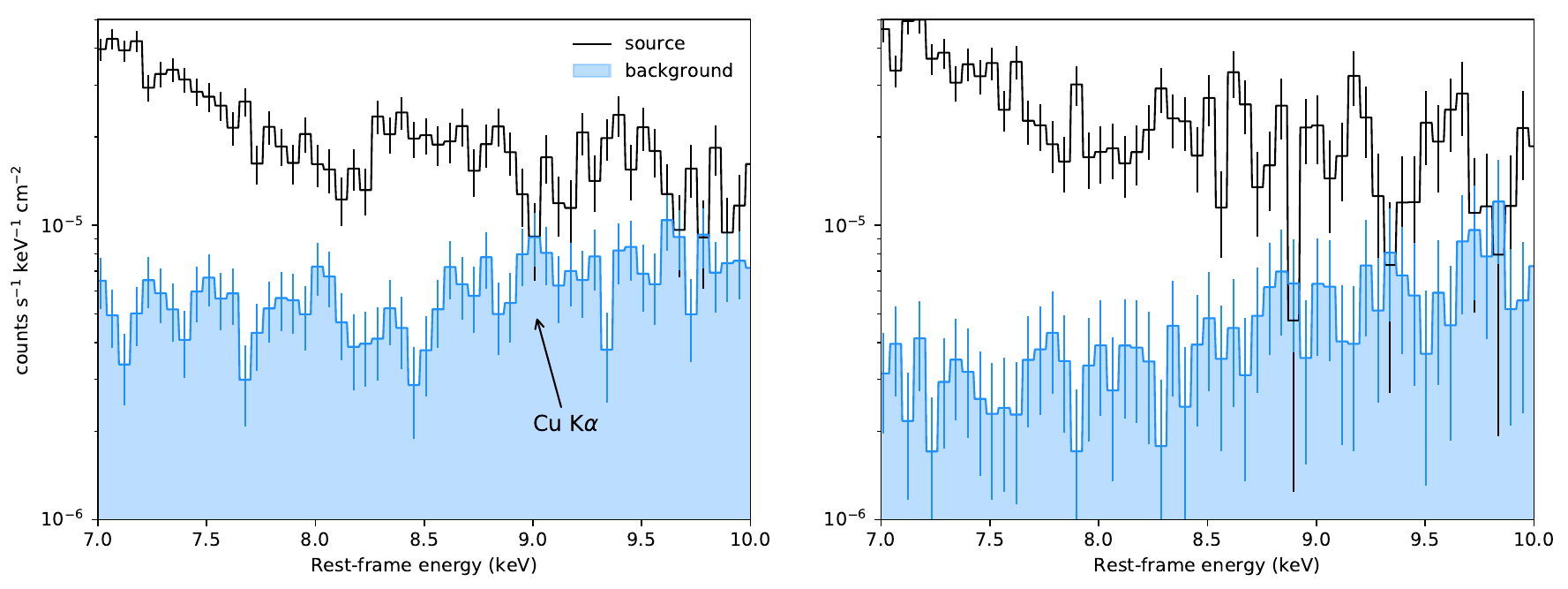}
\caption{High energy source and background spectra for the EPIC-pn (left) and EPIC-MOS (right). The Cu K$\alpha$ line found in the pn background is highlighted, and corresponds to the potential second absorption feature.}
\label{fig_hiresbackground}
\end{figure*}

Firstly, we examine the background spectra at higher resolution in Fig.~\ref{fig_hiresbackground}. For this figure, we bin every ten channels, rather than using the signal-to-noise based binning we used for spectral fitting. The EPIC-pn has a strong Cu line in the internal background, which can potentially produce false absorption features, but there are no strong lines between Fe~K$\alpha$ at 6.4 and Au~L$\alpha$ in the MOS \citep{Lumb02}.

\label{appendix_background}

There is a small peak in the pn background at lower energies, aligned with the main absorption feature, which does not correspond to an instrumental background line. This feature is not strong enough to explain the observed absorption (see Fig.~\ref{fig_background}), but may mildly affect the line shape.

\begin{figure}
\centering
\includegraphics[width=\linewidth]{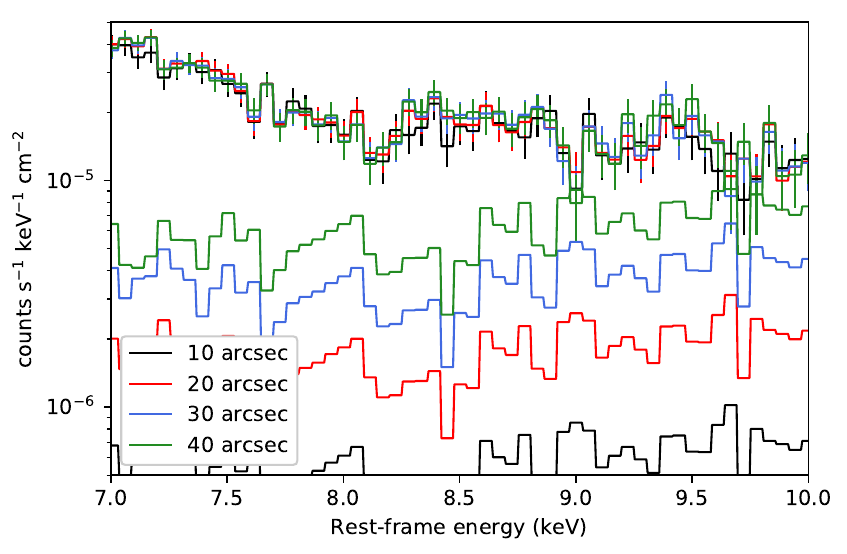}
\caption{Source and background spectra for different source region radii for the EPIC-pn stacked spectrum. For clarity, error bars on the background spectra are not shown.}
\label{fig_pn_sourceradii}
\end{figure}

To further explore the impact of the background region on the shape of the low energy line, we extract source spectra from different circular regions, with radii of 40$^{\prime\prime}$, 30$^{\prime\prime}$, 20$^{\prime\prime}$, and 10$^{\prime\prime}$, centered on the source. This results in different levels of background contribution to the net spectrum, by around an order of magnitude. This has negligible effect on the pn source spectrum (Fig.~\ref{fig_pn_sourceradii}), and only significantly changes one bin at $\sim9.3$~keV in the MOS spectrum, so we conclude that the features discussed here are intrinsic to the source spectrum, not induced by background over- or under-subtraction.

\begin{figure}
\centering
\includegraphics[width=\linewidth]{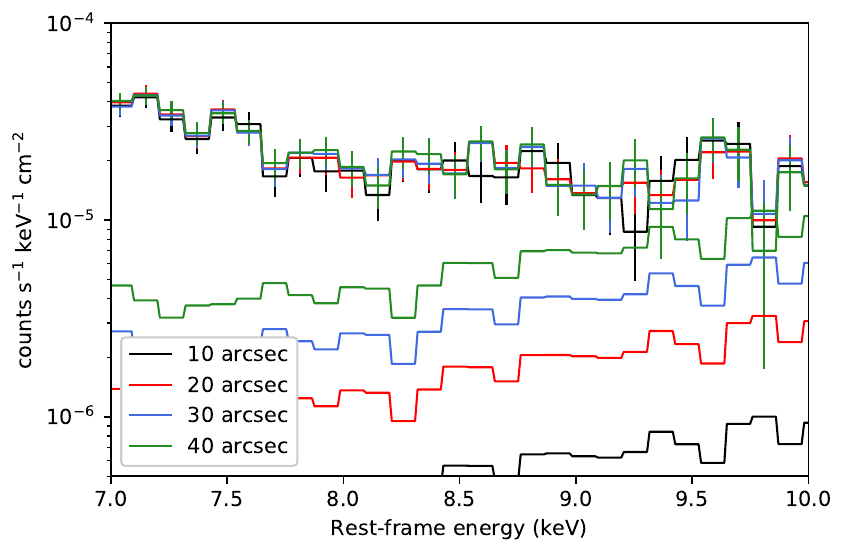}
\caption{Source and background spectra for different source region radii for the EPIC-MOS stacked spectrum. For clarity, error bars on the background spectra are not shown.}
\label{fig_mos_sourceradii}
\end{figure}


\bsp	
\label{lastpage}
\end{document}